\documentclass[journal]{IEEEtran}
\usepackage{amsmath,amsfonts}
\usepackage{algorithmic}
\usepackage{algorithm}
\usepackage{array}
\usepackage[caption=false,font=normalsize,labelfont=sf,textfont=sf]{subfig}
\usepackage{textcomp}
\usepackage{stfloats}
\usepackage{url}
\usepackage{verbatim}
\usepackage{graphicx}
\usepackage{cite}
\usepackage{amsmath,amssymb,amsfonts}
\usepackage{graphicx}
\usepackage{textcomp}
\usepackage{xcolor}
\usepackage{siunitx}
\usepackage{paralist}
\usepackage{booktabs}
\usepackage{makecell} 
\renewcommand{\textcolor}[2]{#2} 
\renewcommand{\color}[1]{}

\def\BibTeX{{\rm B\kern-.05em{\sc i\kern-.025em b}\kern-.08em
    T\kern-.1667em\lower.7ex\hbox{E}\kern-.125emX}}

\begin{document}
%
\title{GEM3D-CIM: \underline{Ge}neral Purpose \underline{M}atrix Computation Using \underline{3D}-Integrated SRAM–eDRAM Hybrid \underline{C}ompute-\underline{I}n-\underline{M}emory-on-Memory Architecture}
%
%
%

\author{Subhradip Chakraborty,~\IEEEmembership{Graduate Student Member,~IEEE,}
        Ankur Singh,~\IEEEmembership{Graduate Student Member,~IEEE,}
        and~Akhilesh R. Jaiswal,~\IEEEmembership{Member,~IEEE}
\thanks{Subhradip Chakraborty, Ankur Singh and Akhilesh Jaiswal are with the Department of Electrical and Computer Engineering, University of Wisconsin Madison, Madison, USA (e-mail: chakrabort42@wisc.edu).}
\thanks{}
\thanks{}}

%
%

\markboth{GEM3D CIM}%
{}
%



\maketitle

\begin{abstract}
With the rapid growth of deep neural networks (DNNs), compute-in-memory (CIM) has emerged as a promising energy-efficient paradigm for accelerating multiply-and-accumulate (MAC) operations. Yet, current CIM architectures are largely limited to dot-product computations and struggle to efficiently support general-purpose matrix operations, such as transpose, element-wise addition, and multiplication. This work presents a 3D-integrated, memory-on-memory SRAM-eDRAM hybrid CIM architecture, implemented in GlobalFoundries 22~nm FDSOI technology, capable of performing general matrix operations directly within the memory crossbar with 4-bit precision. By leveraging a specialized transpose-based architecture, in-memory arithmetic operations, peripheral-aware design, and 3D SRAM--eDRAM integration, the proposed architecture balances latency, energy efficiency, and compute density for general purpose matrix operations while remaining compatible with the conventional CIM dot product architectures. Overall, this memory-on-memory CIM framework generalizes CIM beyond dot products, enabling versatile matrix processing and paving the way for broader applications in AI acceleration and general-purpose high performance computing.

\end{abstract}


\IEEEpeerreviewmaketitle

\section{Introduction}
\IEEEPARstart{T}{he} recent advances in deep learning and neural networks for critical AI applications have led to an increasing demand for data-intensive computations~\cite{dataintensive}. Modern systems suffer from frequent data transfer between processing cores and memory severely limiting both throughput and energy efficiency~\cite{vonneuman}. While transistor scaling predicted by Moore’s law has slowed, the number of processing cores has still steadily increased~\cite{moore}. In contrast, memory architectures, bandwidth, and energy efficiency have not scaled proportionally, giving rise to the well-known memory wall~\cite{memorywall}. To overcome this challenge, a wide range of compute-in-memory (CIM) architectures have been proposed. These architectures primarily target dot-product operations by enabling computation within the memory bit-cells themselves, thereby reducing data movement and improving both energy efficiency and latency in range of workloads including AI~\cite{cim_intro}.

CIM has been explored using various memory technologies, with designs primarily dominated by SRAM, DRAM, and non-volatile memories (NVMs). Prior NVM-based CIM implementations have demonstrated very high compute density and throughput~\cite{nvm1}; however, they often face limitations such as low signal-to-noise ratio (SNR), reduced ON/OFF conductance ratio, high write energy, and fabrication challenges for large-scale integration~\cite{nvm2}. SRAM-~\cite{sramcim} and DRAM-~\cite{dramcim} based analog CIM architectures are generally classified into two categories: charge-domain and current-domain CIM. Charge-domain CIM offers better linearity and higher compute density and energy efficiency but suffers from variations and reduced SNR~\cite{chargeCIM} due to the presence of analog to digital converter (ADC). Current-domain CIM provides improved energy efficiency, compute density, and SNR, but suffers from non-linearity and ADC quantization~\cite{currentCIM}. Despite the massive parallelism inherent in CIM architectures, \textit{performance is ultimately constrained by dot-product reduction}. Specifically, CIM achieves single cycle parallel compute capability only when all rows are activated simultaneously, with each column performing multiply-and-accumulate (MAC) operations whose results must be reduced by ADCs. These characteristics make CIM particularly well-suited for deep learning applications, where MAC operations dominate and high precision is not a critical requirement~\cite{analogcim, mixedprecision}.

While CIM architectures have shown promise in accelerating dot-product operations for inference application in deep learning~\cite{cim11, cim2, cim3}, their applicability to other scientific computing and general-purpose applications remains limited. Domains such as physics-based computation (e.g., Sod shock tube~\cite{sodshock}), recurrent neural networks including Long Short-Term Memory (LSTM) and Gated Recurrent Units (GRU)~\cite{hadamard}, masking operations, and tensor-based algorithms~\cite{tensor} often rely not only on dot products but also on general matrix element-wise operations, such as matrix transpose, element-wise multiplication, and addition. In these applications, both parallelism and precision play a critical role. Conventional CIM accelerators, being optimized primarily for dot products, lose their inherent parallelism advantage when extended to element-wise operations. For example, transpose operations have been demonstrated in CIM frameworks~\cite{transpose1, transpose2}; however, these implementations are typically optimized only for backpropagation, where the dot-product readout enables transpose functionality at the cost of reduced precision. As a result, transpose operations cannot be efficiently stored or represented within the memory crossbar. Similarly, element-wise multiplication, also known as the Hadamard product, has only recently gained attention, with prior works focusing primarily on algorithmic optimization and peripheral circuit design rather than improving the underlying crossbar architecture~\cite{hadamard, hadamardcim}. Matrix addition in CIM has been investigated largely in the context of digital CIM using adder trees~\cite{addtsmc}, but equivalent exploration for elementwise operation in analog CIM frameworks remains limited.

In recent years, advancements in packaging technology and semiconductor manufacturing have enabled the development of monolithic 3D integrated CIM architectures~\cite{cim3d1,cim3d2, cim3d3, cim3d4}. These architectures significantly enhance compute density and open the door for new design possibilities in CIM applications. In this work, we leverage the monolithic 3D-integrated framework to design CIM cores capable of performing \textit{operations beyond conventional dot products}, including matrix transpose, element-wise matrix multiplication, and matrix addition as shown in Fig.~\ref{Overall}. The main contributions of this paper are summarized as follows:

\begin{figure}[!t]
\centering
\includegraphics[width=1\linewidth]{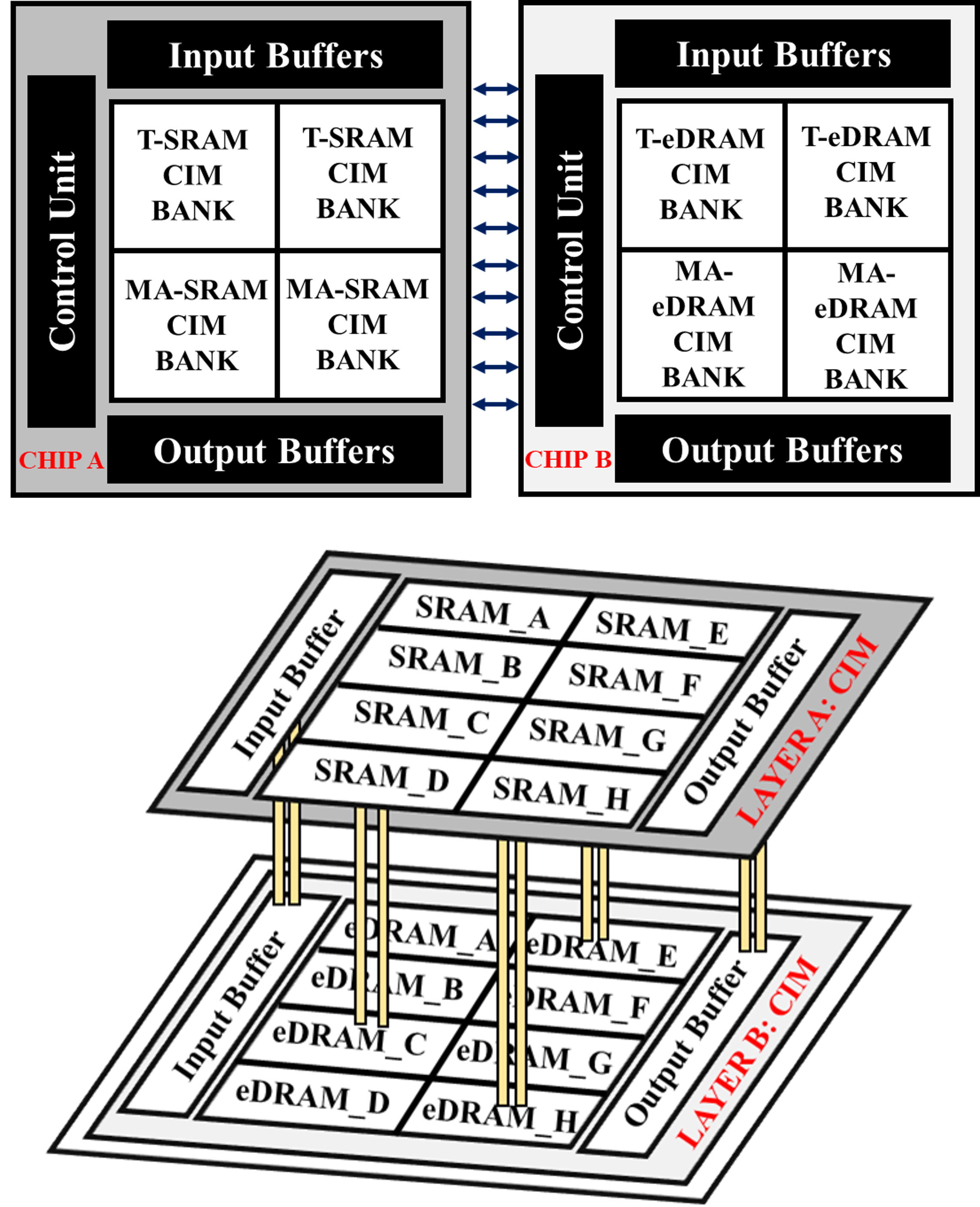}
\caption{The proposed overall architecture comprises two Layers: Layer A, which implements the SRAM array, and Layer B, which implements the eDRAM array. The architecture is organized by monolithic 3D integrating~\cite{imec3d} the two Layers with each one of them divided into subarrays capable of performing matrix transposition, element-wise multiplication, and accumulation operations.}
\label{Overall}
\end{figure}

\begin{itemize}
    \item This work leverages advanced packaging to 3D-integrate SRAM and eDRAM layers, enabling a novel compute-in-memory-on-memory architecture capable of supporting general matrix operations beyond traditional dot-product–centric CIM designs.
    \item Specialized on-memory architecture blocks are introduced to efficiently perform matrix transpose operations within the memory fabric, eliminating costly data movement.
    \item The design further supports element-wise matrix multiplication and accumulation directly inside the memory crossbar, with all core computations executed in the analog domain.
    \item Despite its analog compute nature, both input matrices and computed outputs are represented and stored in fully digital form within the memory crossbar. This ensures seamless integration with conventional digital systems and accelerators.
    \item The architecture repurposes 8T SRAM cells as DACs and employs eDRAM-based structures as lightweight ADCs, minimizing peripheral overhead without sacrificing precision.
\end{itemize}

The remainder of this paper is organized as follows. Section~II introduces the proposed memory bit-cell designs. Section~III describes the matrix transpose operation, while Section~IV presents the element-wise matrix multiplication and addition operations. Section~V discusses the conventional MAC operation. Section~VI provides the results obtained using the GlobalFoundries 22\,nm PDK, and Section~VII concludes the paper.


\begin{figure*}[!t]
\centering
\includegraphics[width=1\linewidth]{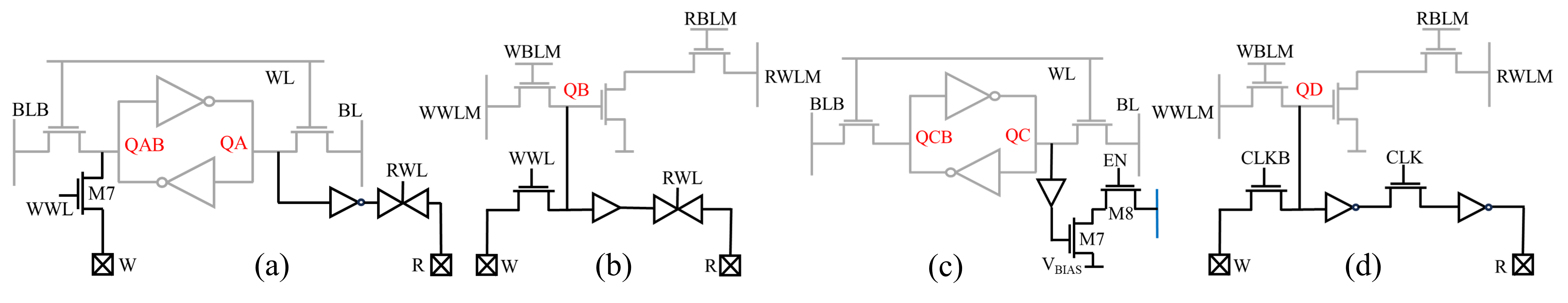}
\caption{Transpose based: (a) T-SRAM bit-cell; (b) T-eDRAM bit-cell; Multiplication and Addition based: (c) MA-SRAM bit-cell; (d) MA-eDRAM bit-cell. }
\label{bitcells}
\end{figure*}

\section{Proposed Memory Bit-cells}

The proposed 3D-integrated CIM architecture for general-purpose matrix operations is realized using two vertically stacked device Layers (Layer A and Layer B) formed by a monolithic 3D fabrication flow, with fine-pitch monolithic inter-Layer vias providing high-density vertical electrical connections between Layers~\cite{monopro1, monopro2, monopro3}. Layer A contains SRAM arrays supporting 4-bit word operations, while Layer B consists of eDRAM arrays with 4/8-bit word width, enabling efficient matrix transpose, elementwise multiplication, and addition functionalities. This integration technique leverages the high speed and stability of SRAM with the density and energy efficiency of eDRAM, providing a balanced platform for matrix-centric computation.

\subsection{Transposable SRAM (T-SRAM) Bit-cell}

The proposed bit-cell extends the conventional 6T SRAM structure with an additional 5T circuitry to enable transpose functionality as shown in Fig.~\ref{bitcells}(a). The cell operates in two distinct modes: (1) SRAM Mode, where it functions as a standard 6T SRAM supporting conventional read/write operations and (2) Transpose (T) Mode, which leverages dedicated wordline controls (WWL and RWL) to perform simultaneous read and write operations across different bit-cells without disturbing the stored values during read. In T mode, when the RWL is asserted, data is sensed from the QA node and propagated through an inverter to the R node. Importantly, because QA directly drives the inverter’s input, the stored value at QA remains unaffected during readout. To ensure balanced signal transitions for both logic states, the RWL logic is implemented using a transmission gate (TG), providing symmetric rise and fall times essential for reliable transpose operations under PVT variations. Similarly, when WWL is asserted, the value at node W is written onto the QAB node of the 6T SRAM, thereby enabling write operations in T mode. This dual-mode functionality facilitates efficient matrix transpose operations, described in detail in the next section while preserving the conventional benefits of SRAM storage.

\subsection{Transposable eDRAM (T-eDRAM) Bit-cell}

The proposed eDRAM bit-cell consists of a conventional 3T eDRAM core with an additional 7T circuitry to support read and write operations in transpose (T) mode. Similar to the T-SRAM, the T-eDRAM operates in two modes: as a conventional eDRAM or in T mode for transpose functionality. Unlike SRAM, the eDRAM stores data on a single node (QB) without a complementary pair, necessitating the use of a buffer to provide isolation. This ensures that the stored charge at QB remains undisturbed when RWL is asserted, while simultaneously driving the output at node R, as shown in Fig.~\ref{bitcells}(b). In T-SRAM, isolation can be achieved with a single inverter since data is read from QA and written into QAB of a different bit-cell. \textcolor{red}{However, in T-eDRAM, both the read and write operations are performed through QB, requiring a buffer for correct data write. During write operations, when WWL is asserted, the data at node W is transferred to the QB node, thereby enabling write functionality in T mode.}

\subsection{Multiplication and Addition Enabled SRAM (MA-SRAM) Bit-Cell}

Our approach for implementing element-wise multiplication and addition leverages an 8T SRAM bit-cell. The additional two transistors in the 8T design are utilized to realize a digital-to-analog converter (DAC)~\cite{8tdac}. For a 4-bit word, the SRAM operating at 0.8 V does not provide sufficient sensing margin to reliably distinguish 16 levels. To address this, an auxiliary buffer circuit, shown in Fig.~\ref{bitcells}(c), is activated to generate a higher analog sensing voltage. The buffer, as well as transistors M7 and M8, are implemented using devices that support
a supply voltage of 1.8\,V. Accordingly, the \textit{EN} signal is overdriven to 1.8\,V, while
the bias voltage $V_{\mathrm{BIAS}}$ is set to 1.2\,V. These values are used for simulation with the current PDK. However, they can be scaled down by employing ultra low-threshold voltage transistors, thereby enabling operation at lower supply levels and increasing the analog sensing voltage margin~\cite{ultralowtx}. In the MA-SRAM bit-cell, the two extra transistors (M7 and M8) are sized with varying widths across the 4-bit word in the ratio 8:4:2:1, thereby producing the MSB-to-LSB weighting effect. Each MA-SRAM bit-cell contributes a current proportional to the value stored at node QC of the 6T SRAM, and these weighted currents are summed in the current domain. The aggregated current is then passed through a parallel transistor network to generate an equivalent analog voltage.

\subsection{Multiplication and Addition Enabled eDRAM (MA-eDRAM) Bit-Cell}

The MA-eDRAM bit-cell consists of 9 transistors, as illustrated in Fig.~\ref{bitcells}(d), and is organized into an 8-bit word. The bit-cells within each word are arranged to function as an 8-bit linear feedback shift register (LFSR) counter~\cite{lfsrcounter}, thereby generating 64 distinct output levels, as explained in detail in subsequent sections. This 8-bit eDRAM word is used as a memory storage but also is utilized to realize an ADC, with the detailed operation described in the following section. For the ADC operation, the MA-eDRAM bit-cell structure operates in a clocked manner, where the positive phase of the clock is used to read the data value QD from the 3T eDRAM, while the negative phase is used to write the corresponding value into the subsequent eDRAM bit-cell.

\begin{figure*}[!t]
\centering
\includegraphics[width=1\linewidth]{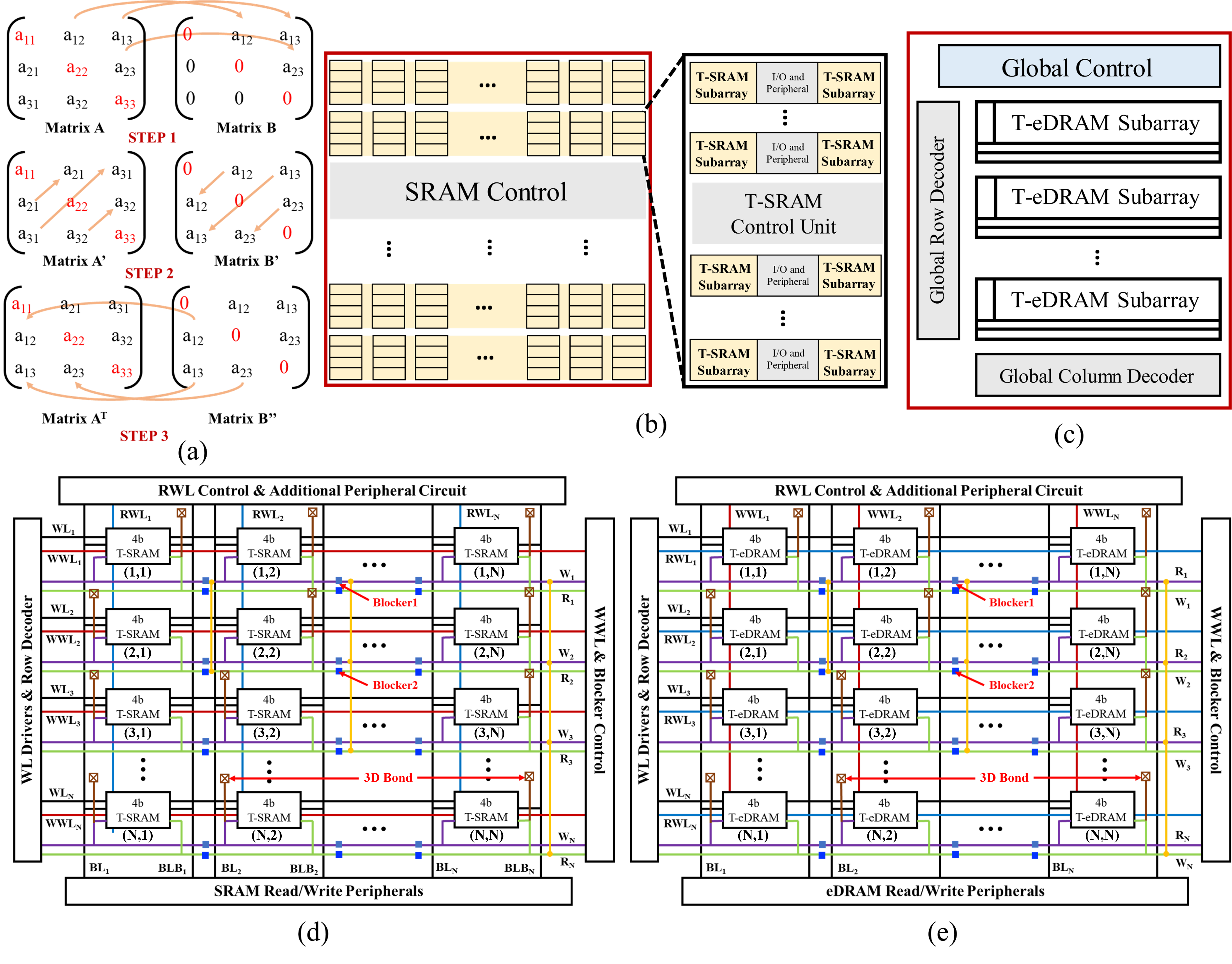}
\caption{
(a) Illustration of a $3 \times 3$ matrix transpose operation, demonstrating the reordering of matrix elements along the diagonal; 
(b) SRAM multi–sub-array architecture comprising both Transpose-SRAM (T-SRAM) and Multiply–Accumulate SRAM (MA-SRAM) banks; 
(c) Sub-array level organization of the T-eDRAM architecture; 
(d) Proposed T-SRAM crossbar implementing a 4-bit word; and 
(e) Proposed T-eDRAM crossbar with a 4-bit word structure, optimized for multi-bit transpose operation.
}
\label{T_array}
\end{figure*}

\section{Matrix Transpose}

The matrix transpose operation can be broadly divided into three steps. Consider a $3 \times 3$ square matrix, where the diagonal elements remain unchanged during transpose operation, while the elements above and below the diagonal exchange their positions to ensure that $a_{ij} = a_{ji}$. The original matrix is stored in Layer A (Matrix A), and the process begins by copying the upper-diagonal elements of the matrix into their corresponding positions in the upper diagonal of Layer B (Matrix B), as illustrated in Fig.~\ref{T_array}(a). In the next step, an internal swap is performed within both Layers simultaneously: in Layer A, the lower-diagonal elements are copied into the corresponding upper-diagonal positions (for instance, $a_{21}$ is copied into $a_{12}$ and $a_{31}$ into $a_{13}$), while in Layer B, the reverse operation is carried out (copying from upper-diagonal elements to lower-diagonal elements). In the final step, the lower-diagonal elements stored in Layer B are copied back into their respective lower-diagonal positions in Layer A, thereby completing the transpose operation while preserving the diagonal elements of the original matrix as detailed in Algorithm 1.

\subsection{Layer A: T-SRAM Sub-Array}

We demonstrate the matrix transpose operation using the T-SRAM bit-cell described in the previous section, configured for a bit-precision of 4 bits with dedicated sub-arrays designed exclusively for transpose operations. The memory crossbar architecture, shown in Fig.~\ref{T_array}(d), is scalable with $N$ with 4-bit word size, similar to a conventional memory array. In this design, the read word-lines (RWL) are routed vertically and the write word-lines (WWL) are routed horizontally, while the remaining routing follows the conventional SRAM layout.The read (R) shown in "green" and write (W) lines shown in "purple" are routed horizontally, and additionally the R and W lines are connected among each other, as shown using a "yellow" line in Fig.~\ref{T_array}(d). To enable the transpose operation, additional blockers in the form of transmission-gate (TG) switches are introduced between adjacent memory rows, with each bit-cell incorporating a single 3D bond. 

The original matrix is stored in Layer A. In the first step of the transpose operation, the values from the upper diagonal of Layer A are read by activating all the RWL lines in parallel while switching off all the blockers (Blocker 1 and 2) marked in Fig.~\ref{T_array}(d) along the horizontal lines connected to the read and write paths. Switching off the blockers is essential to ensure that all data can be read simultaneously and distinctly across the whole sub-array. To transfer the upper-diagonal matrix elements from Layer~A to Layer~B, the $R$ nodes of the corresponding rows in Layer~A are vertically connected to the $W$ nodes of the sub-array in Layer~B via 3D connections. During this operation, the RWL lines associated with the upper-diagonal entries in Layer~A are activated, while the matching WWL lines in Layer~B are simultaneously enabled, allowing the data to copy from Layer A to Layer B. For example, in a $3{\times}3$ matrix, the upper-diagonal elements $(1,2)$, $(1,3)$, and $(2,3)$ are selected by turning on their respective RWLs in Layer~A and the corresponding WWLs in Layer~B, with each $R$ node of these elements in Layer~A linked to its paired $W$ node in Layer~B through the monolithic 3D interconnects.

\begin{algorithm}[H]
\caption{Matrix Transpose Operation Flow Across 3D-Integrated Layers A and B}
\begin{algorithmic}[1]
\STATE \textbf{Load Matrix A into Layer A:}  
Load the input square matrix (with proper padding) into the SRAM arrays of Layer A, initializing the memory banks for subsequent operations.

\STATE \textbf{Transfer Operand from Layer A to Layer B:}  
Move the upper diagonal matrix data from Layer A to Layer B through the 3D interconnect.

\STATE \textbf{Perform Internal Shifting in Layer A and B:}  
Within Layer A and B, execute the internal shifting (copying data from lower diagonal to upper diagonal in Layer A and vice-versa in Layer B) operations required for matrix transpose.

\STATE \textbf{Transfer Operand from Layer B back to Layer A:}  
Return the processed lower diagonal data from Layer B to Layer A through the 3D interconnect, allowing completion of the matrix transpose operation.

\STATE \textbf{Read Result from Layer A:}  
Read the final computed matrix result from Layer A, completing the end-to-end in-memory transpose workflow.
\end{algorithmic}
\end{algorithm}

In the second step, the lower-diagonal values are internally copied into their corresponding upper-diagonal positions, performed one column at a time and thus requiring $N-1$ cycles. During this process, the RWL and WWL (overdriven) lines are activated sequentially, while Blocker 1 is switched off and Blocker 2 is switched on. For illustration, consider a $3 \times 3$ matrix array. When $RWL_1$ and $WWL_1$  are activated, the data stored at positions $(2,1)$ and $(3,1)$ are read and simultaneously written to positions $(1,2)$ and $(1,3)$, respectively, through the R and W lines. Specifically, the data at $(2,1)$ is read by $R_2$ and written only to $(1,2)$ via $W_1$, which is connected to $R_2$ through the ``yellow'' wire shown in Fig.~\ref{T_array}(d). At the same time, Blocker 1 ensures that the data is not written to $(1,1)$ or $(1,3)$. In the following cycle, when $RWL_2$ and $WWL_2$ are activated, the data at $(3,2)$ is copied into $(2,3)$, thereby completing the second step of the operation.

In the third and final step, the WWL lines (overdriven) are activated while keeping both Blocker 1 and Blocker 2 in the OFF state, thereby enabling the write operation. In this phase, the $W$ nodes of the lower diagonal in Layer A are connected to the $R$ nodes in Layer B, allowing the lower-diagonal data from Layer B to be written back into their corresponding locations in Layer A. This completes the transpose operation in $N+1$ cycles. As discussed earlier, the WWL lines are intentionally overdriven during the write phase to ensure correct data programming across all T-SRAM bit-cells in a large array. Recall that, as described in Step~2 (and detailed further in the next subsection), Layer B undergoes a similar in-memory copy operation. As a result, Layer B holds the transposed data in its lower-diagonal elements. Thus, the original matrix is first loaded into Layer A, and upon completion of the full operation, Layer A now stores the transpose of the original matrix.

\subsection{Layer B: T-eDRAM Sub-Array}

We perform a similar operation to the one described in the previous subsection, this time utilizing Layer B, which consists of T-eDRAM bit-cells. The transpose operation in Layer B is carried out in three steps, with each word being 4-bits wide and supported by a dedicated subarray designed specifically for this purpose. To ensure uniformity, the memory crossbar architecture in Layer B is configured to match the capacity of the T-SRAM subarrays in Layer A. In Layer B, the RWL lines are routed horizontally, while the WWL lines are routed vertically, with the remaining interconnections following the conventional 3T eDRAM routing structure. The R and W lines highlighted with "purple" and "green", respectively are routed horizontally and connected to each other via "yellow" wires shown in Fig.~\ref{T_array}(e).

In the first step, the upper diagonal elements from the T-SRAM subarray in Layer A are read, and their read ports (R) are connected to the write nodes (W) of the upper diagonal elements of T-eDRAM bit-cells in Layer B through 3D bonds located along the upper diagonal. This enables data to be copied from Layer A to Layer B in a single cycle, since all elements are operated in parallel by simultaneously activating all WWL lines. The second step involves an internal in-memory copy within the T-eDRAM subarray of Layer B, where the upper diagonal elements are transferred to their corresponding lower diagonal positions. This is achieved by sequentially activating pairs of RWL and WWL lines (overdriven), $RWL_1$ with $WWL_1$, $RWL_2$ with $WWL_2$, and so forth, up to $RWL_{N-1}$ with $WWL_{N-1}$. In the final step, all RWL lines are activated, and the R nodes of the lower diagonal elements in the T-eDRAM subarray are connected to the W nodes of the T-SRAM subarray in Layer A via 3D bond. This transfers the lower diagonal values from Layer B back to Layer A, thereby completing the transpose operation. Overall, the procedure requires only $N+1$ cycles: one cycle for the first step,
$N-1$ cycles for the second step, and one cycle for the third step. In comparison, a conventional transpose implemented using sequential read write operations typically requires $2N$ cycles.


\begin{figure}[!t]
\centering
\includegraphics[width=1\linewidth]{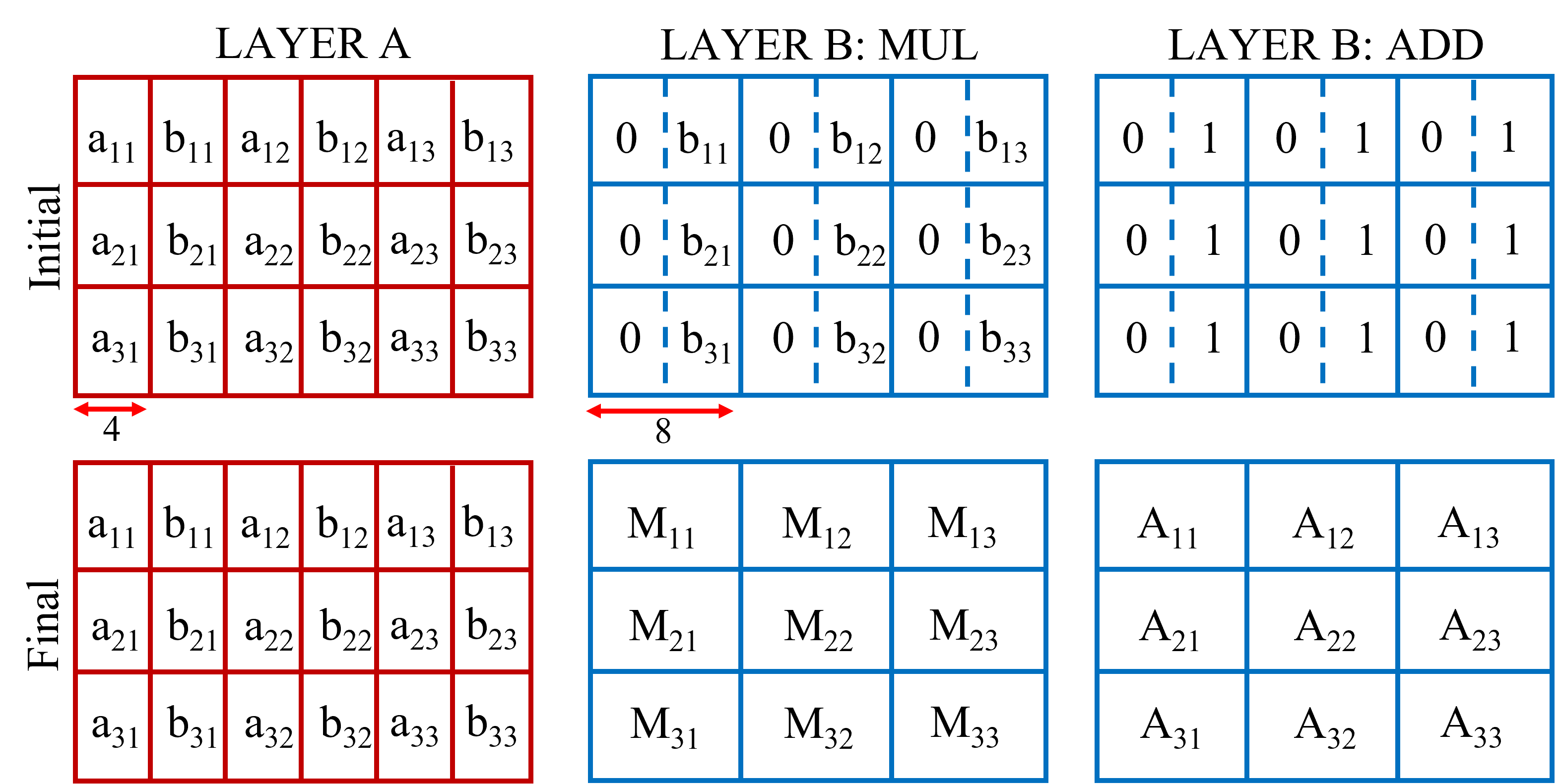}
\caption{\textcolor{blue}{Mapping of elements of Matrix A and B to perform element-wise matrix multiplication and addition operations.}}
\label{Mapping_MA}
\end{figure}

\begin{figure*}[!t]
\centering
\includegraphics[width=1\linewidth]{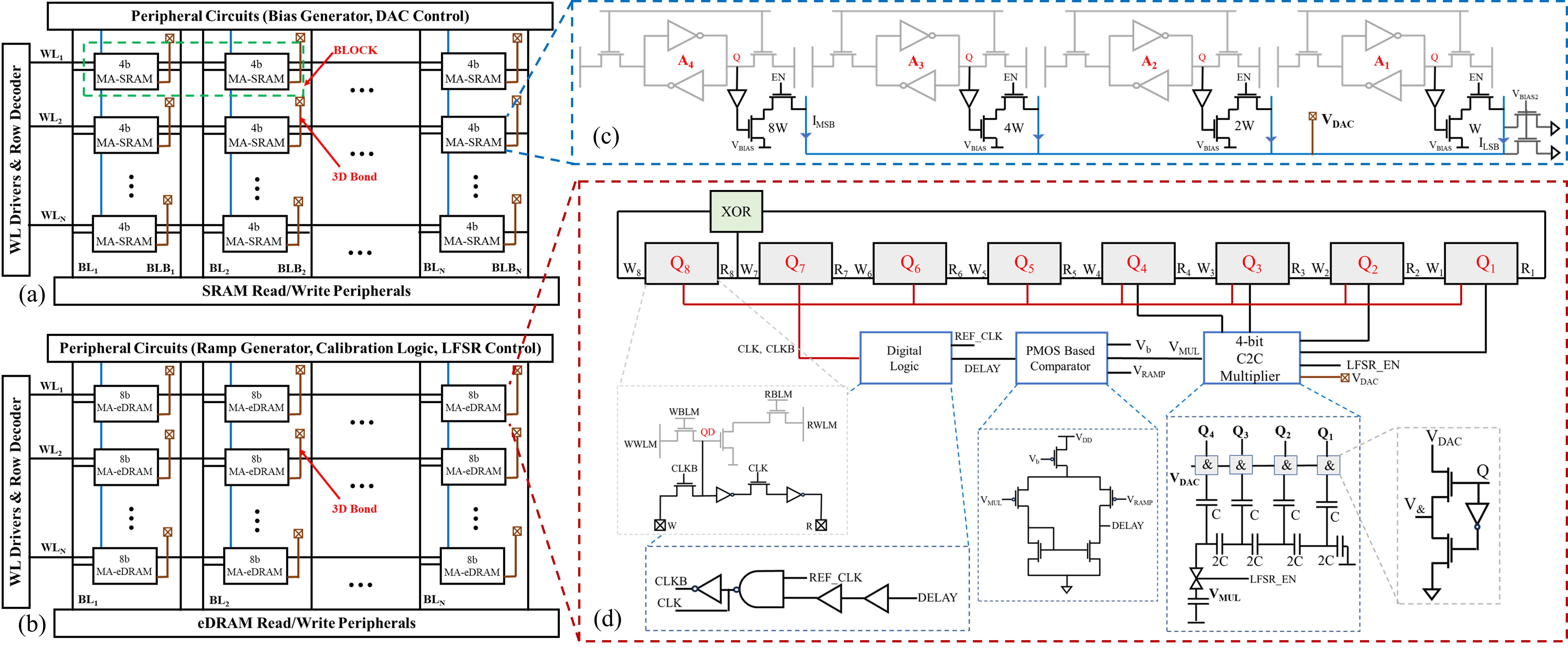}
\caption{
(a) Proposed MA-SRAM sub-array implementing a 4-bit word structure; 
(b) Proposed MA-eDRAM sub-array designed with an 8-bit word organization; 
(c) Digital-to-analog conversion (DAC) realized using MA-SRAM bit-cells; and 
(d) 8-bit LFSR-based MA-eDRAM architecture capable of performing analog multiplication, followed by analog-to-digital conversion.
}

\label{M_array}
\end{figure*}

\section{Multiplication/Addition Operation}

We detail the procedure used to perform the element-wise multiplication and addition operations within the memory crossbar. 
For the element-wise multiplication or addition of two matrices, A and B, the matrices are stored in Layer A, which consists of MA-SRAM bit-cells as described in the section II. Each word is 4-bit wide, and the corresponding elements of matrices A and B are stored next to each other, as shown in Fig.~\ref{Mapping_MA}, thereby forming a combined 8-bit block. The analog values of matrices A and B are generated using a current-based DAC (Fig.~\ref{M_array}(c)). For the addition operation, both the elements of matrices A and B are converted to analog form, whereas for multiplication only the analog value of A is generated. Layer B, which consists of MA-eDRAM bit-cells (each word being 8-bit wide), is reconfigured to perform analog-to-digital conversion. The two 4-bit values corresponding to elements of A and B are multiplied or added, producing an equivalent 6-bit digital output. This output corresponds to the 8-bit LFSR codes.

\subsection{MA-SRAM Sub-array}

The MA-SRAM sub-array consists of 4-bit words, with each word representing an element in a matrix. 
The values stored in the SRAM are amplified through a 1.8~V buffer to provide a higher signal margin for the DAC. The DAC circuit is constructed using MA-SRAM bit-cells arranged in an MSB-to-LSB pattern by appropriately weighting the access transistors, as shown in Fig.~\ref{M_array}(c). When a stored bit is "1", a current is drawn from a bias voltage ($V_{BIAS}$) of 1.2~V; otherwise, no current flows when the stored bit is "0". These currents are summed and passed through a parallel combination of transistors that convert the equivalent current into an analog voltage. This analog voltage is then supplied to Layer B through monolithic 3D interconnect, which consists of MA-eDRAM bit-cells, to further perform the remaining steps for the element-wise multiplication operations. For analog addition the currents from neighbouring words are summed up and then converted to voltage and passed on to Layer B for analog to digital conversion as shown in Fig.~\ref{Addition}.

\begin{algorithm}[h]
\caption{Matrix Element-wise Multiplication/Addition Operation Flow Across 3D-Integrated Layers A and B}
\begin{algorithmic}[1]

\STATE \textbf{Load Matrices into Layers:}  
Load Matrix A and B into both Layer A, and load Matrix B into Layer B (only for multiplication), initializing memory banks for computation.

\STATE \textbf{DAC Operation:}  
Perform 4-bit DAC operation to generate the analog values corresponding to elements of Matrix A and B.

\STATE \textbf{Transfer VDAC from Layer A to Layer B:}  
Move the generated analog voltage (VDAC) from Layer A to Layer B through the 3D interconnect (only elements of matrix A during multiplication and elements of both A and B during addition).

\STATE \textbf{Perform Multiplication/Addition in Layer B:}  
Execute the analog element-wise multiplication/addition operation on Layer B using the received VDAC and stored Matrix B values.

\STATE \textbf{eDRAM Write for LFSR Start Bits:}  
Write the initial values for the LFSR into eDRAM to prepare for analog-to-digital conversion.

\STATE \textbf{LFSR-Based Analog-to-Digital Conversion:}  
Use the LFSR mechanism to convert the analog multiplication/addition results into digital format.

\STATE \textbf{Read Result from Layer B:}  
Read the final computed results from Layer B, completing the analog in-memory multiplication/addition workflow.

\end{algorithmic}
\end{algorithm}

\subsection{MA-eDRAM Subarray}

The MA-eDRAM subarray consists of 8-bit words along with additional computing circuitry to perform element-wise matrix multiplication and addition operations. 
Consider the multiplication case where we want to perform element-wise multiplication between two matrices A and B, with each element being 4-bit wide. 
The elements of matrix A and B are stored in the MA-SRAM subarray in Layer A, while the elements of matrix B are also stored in Layer B in the LSB 4 bits of the 8-bit word. A capacitive C2C multiplier is employed between the 4-bit wide elements of matrix B and the equivalent analog voltages of the 4-bit elements of matrix A, which are generated by the DAC ($V_{DAC}$) in MA-SRAM subarray in Layer A as shown in Fig.~\ref{M_array}(d). The output of the multiplier ($V_{MUL}$) is passed to a PMOS-based differential amplifier, which converts the analog multiplication output voltage into a time-delayed signal. The delay is generated by comparing the analog voltage produced by the multiplier with a globally shared ramp generator. If the analog value is greater than the ramp signal, the output of the comparator acts as the delayed signal which is further passed through a digital logic block, and an equivalent CLK and CLKB signal are generated based on the delay and a reference clock. This reference clock serves as the clock signal for the 8-bit LFSR-based eDRAM counter. 

After the analog multiplication is completed, the LFSR enable signal (LFSR\_EN) samples the analog multiplication voltage, and the 8-bit MA-eDRAM word is initialized to its starting position for counting, i.e., 00000001 (the LFSR starting point). 
The read and write nodes of consecutive MA-eDRAM bit-cells in a word are connected to facilitate the in-memory shifting operation for positions Q7 to Q1, whereas Q8 is generated by XORing Q7 and Q1 (the LFSR logic). 
Depending on the number of CLK pulses generated (which is equivalent to the delay), the eDRAM-based LFSR counter increments and converts the analog multiplicative value into its 8-bit LFSR code, Finally, the output of the computation is naturally available in its 8-bit LFSR-encoded form. If required, this 8-bit LFSR code can be converted into the corresponding 6-bit digital value using a lookup table (LUT) during a subsequent read operation. The overall step-by-step procedure is summarized in Algorithm~2.

\begin{figure}[!t]
\centering
\includegraphics[width=1\linewidth]{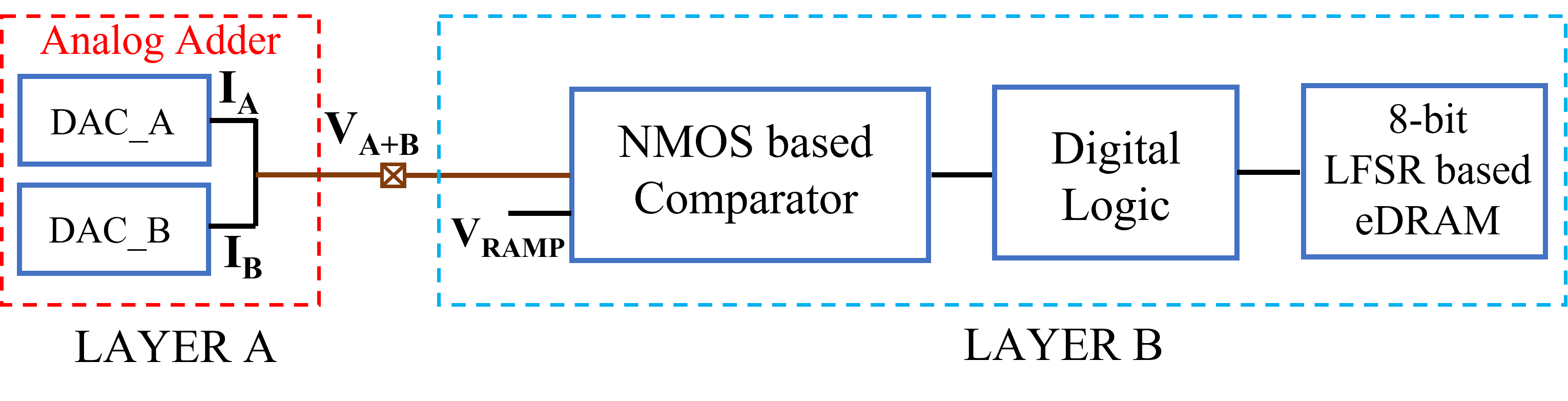}
\caption{
\textcolor{blue}{Data flow for performing analog addition using DAC\_A and DAC\_B, each corresponding to a 4-bit MA-SRAM word, followed by analog-to-digital conversion within an 8-bit MA-eDRAM word.}
}
\label{Addition}
\end{figure}

Consider the element-wise addition operation, where almost the same 8-bit MA-eDRAM word is used in the subarray dedicated for addition operation. 
Instead of employing the C2C multiplier, an analog addition in current domain is performed in Layer A as shown in Fig.~\ref{Addition} and the analog output is passed through a NMOS based comparator followed by the digital logic and LFSR based eDRAM. 
Both matrices A and B consist of 4-bit elements that are stored in MA-SRAM subarray in Layer A, and their analog values are generated by the DAC in the MA-SRAM word. These analog signals are then transmitted to the 8-bit words in Layer B through 3D monolithic interconnect.  The remaining operations proceed in the same manner as in the multiplication case: the equivalent delay is generated based on the analog voltage, and the LFSR counter increments according to the number of CLK pulses.

\section{MAC operations}

The conventional MAC operations can be implemented using the MA-SRAM bit-cell presented in Layer A, which is based on an 8T SRAM structure. The weights are stored in the 6T SRAM portion, while the enable signal (EN) corresponds to the input activation (IA) shared across each row. The 4-bit word follows the same structure as the MA-SRAM word, employing weighted transistors to represent the MSB-to-LSB scaling effect as described earlier. The outputs from each word are connected along a column to perform accumulation in the current domain. The final accumulated output can either be routed to a dedicated ADC for high-precision conversion or sent to Layer B for analog-to-digital conversion using the LFSR-based eDRAM mechanism. Thus, in addition to supporting the elementary matrix operations described earlier, the proposed array architecture is also capable of performing conventional dot-product computations.

\section{Results and Discussion}

\begin{figure}[!t]
\centering
\includegraphics[width=1\linewidth]{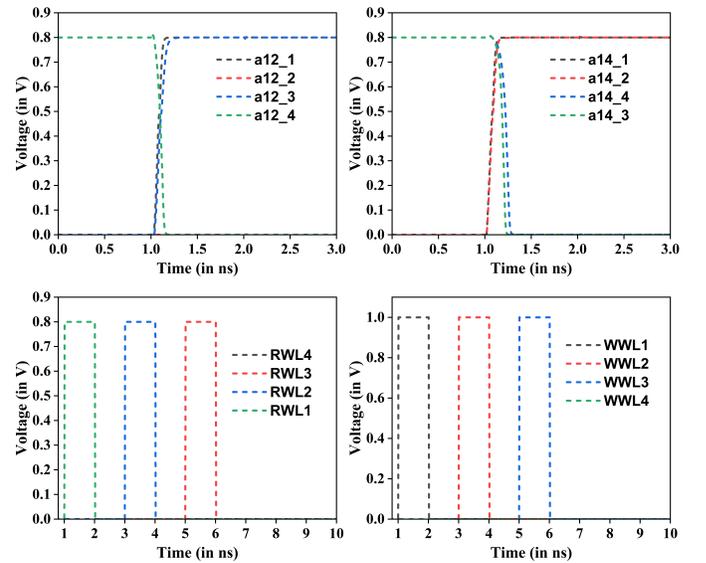}
\caption{
\textcolor{blue}{Transient simulation on GF~22\,nm FDSOI technology demonstrating the copying of lower-diagonal elements to the upper-diagonal in a T-SRAM array during a $4 \times 4$ matrix transpose operation. 
In this example, $a_{21} = 0101$, $a_{41} = 0011$, while $a_{12}$ switches from $1000$ to $0101$ (corresponding to $a_{21}$) and $a_{14}$ switches from $1100$ to $0011$ (corresponding to $a_{41}$).}
}
\label{SRAM_T_Transient}
\end{figure}

\begin{figure}[!t]
\centering
\includegraphics[width=1\linewidth]{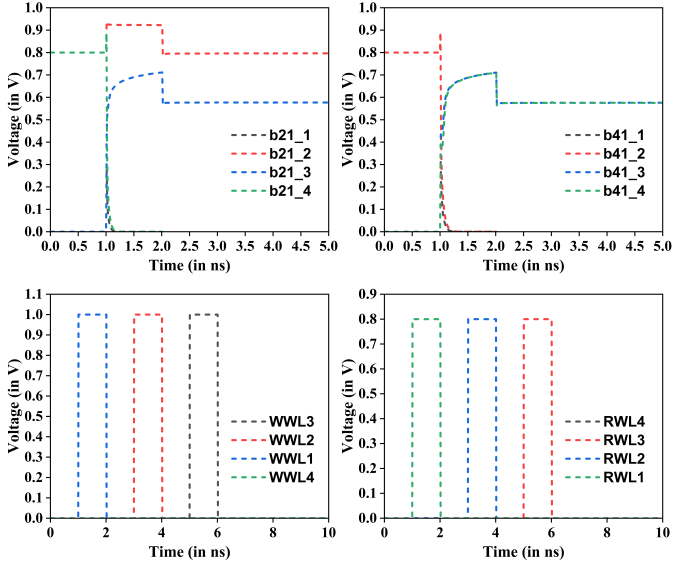}
\caption{
\textcolor{blue}{Transient simulation demonstrating the copying of upper-diagonal elements to the lower-diagonal in a T-eDRAM array during a $4 \times 4$ matrix transpose operation. 
In this example, $b_{12} = 0110$, $b_{14} = 1100$, while $b_{21}$ switches from $1011$ to $0110$ (corresponding to $b_{12}$) and $b_{41}$ switches from $0011$ to $1100$ (corresponding to $b_{14}$).}
}
\label{eDRAM_T_Transient}
\end{figure}

\begin{figure}[!t]
\centering
\includegraphics[width=1\linewidth]{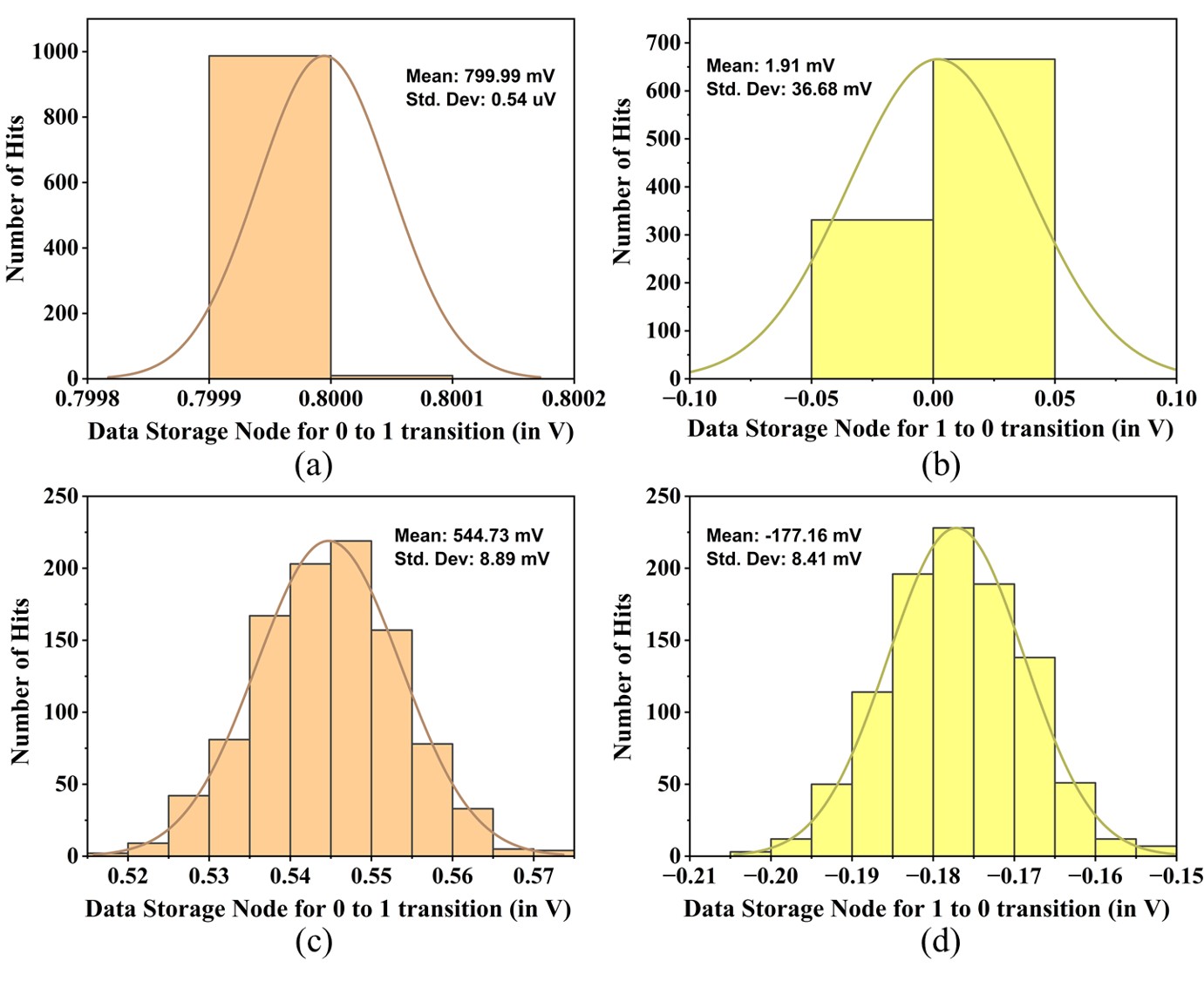}
\vspace{-0.2 in}
\caption{
Variation-aware transient simulations after write operation showing: 
(a) 0-to-1 transition in T-SRAM; 
(b) 1-to-0 transition in T-SRAM; 
(c) 0-to-1 transition in T-eDRAM; and 
(d) 1-to-0 transition in T-eDRAM.
}
\label{T_MC}
\end{figure}
\subsection{Transpose Operation}

We implement the transpose operation with a supply voltage ($V_{DD}$) of 0.8~V, an RWL voltage of 0.8~V, and an overdriven WWL voltage of 1~V in both the T-SRAM and T-eDRAM subarrays. To validate the proposed scheme, we perform an example case of a $4 \times 4$ matrix transpose operation, where each matrix element is 4-bit wide. During this computation, four rows are activated and sixteen columns are accessed within the transpose memory subarray. The transient simulation results for Step~2 (as described in the earlier section), where the lower diagonal elements are copied into the upper diagonal positions for the T-SRAM and the opposite occurs for the T-eDRAM subarray, are shown in Fig.~\ref{SRAM_T_Transient} and Fig.~\ref{eDRAM_T_Transient}, respectively. \textcolor{blue}{Fig.~\ref{SRAM_T_Transient} illustrates the swapping operation in the T-SRAM, where the elements $a_{21}$ and $a_{41}$ are read and written into $a_{12}$ and $a_{14}$, respectively. Similarly, Fig.~\ref{eDRAM_T_Transient} depicts the T-eDRAM operation, where the elements $b_{12}$ and $b_{14}$ are read and written into $b_{21}$ and $b_{41}$, respectively.} To further validate the correctness of the transpose operation, we perform Monte Carlo simulations with 1000 samples. Fig.~\ref{T_MC}(a) and (b) show the voltage switching characteristics (0~→~1 and 1~→~0 transitions) for the T-SRAM, while Fig.~\ref{T_MC}(c) and (d) present the corresponding histogram plots for the T-eDRAM during the transpose operation.

\begin{figure}[!t]
\centering
\includegraphics[width=1\linewidth]{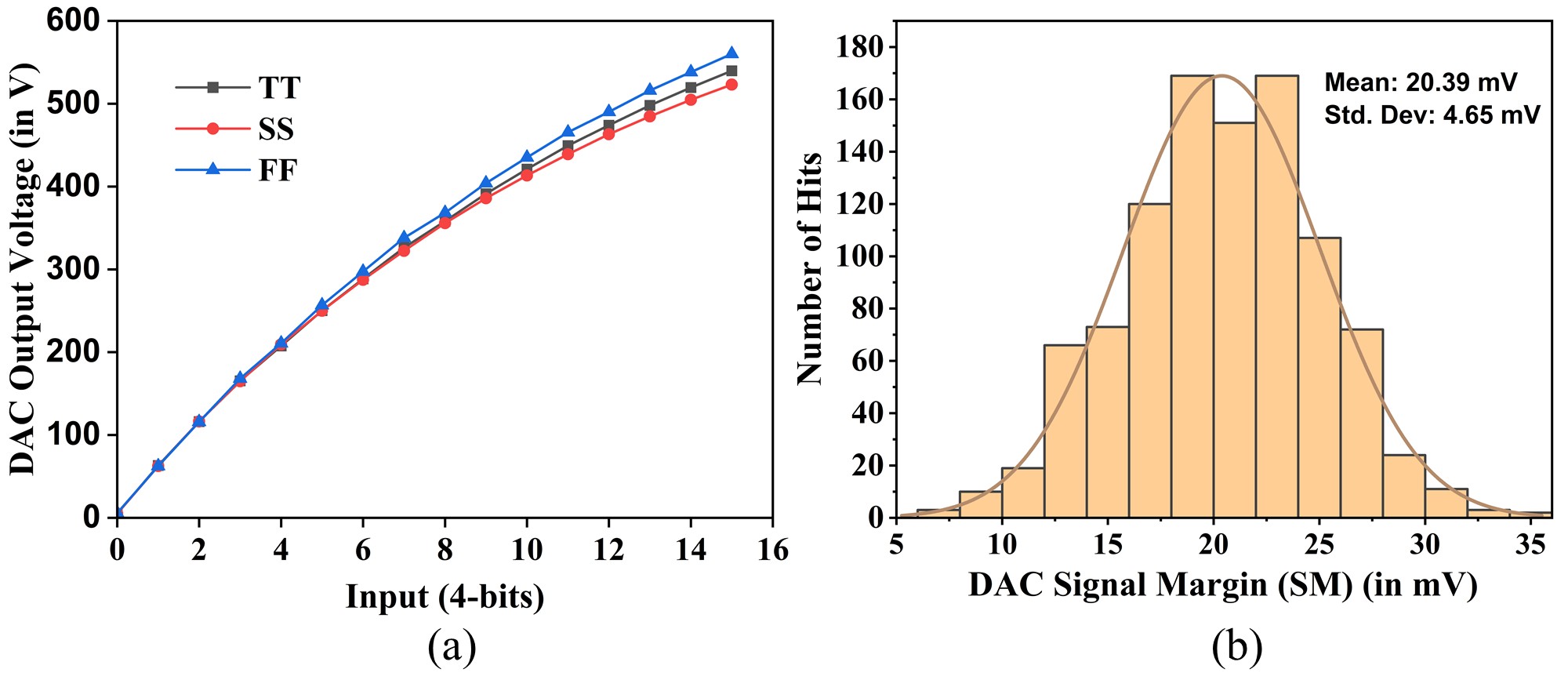}
\vspace{-0.2 in}
\caption{
(a) T-SRAM DAC output voltage as a function of 4-bit digital input across different process corners; 
(b) Variation of signal margin for the 4-bit DAC.
}
\label{DAC_PC_MC}
\end{figure}

\begin{figure}[!t]
\centering
\includegraphics[width=1\linewidth]{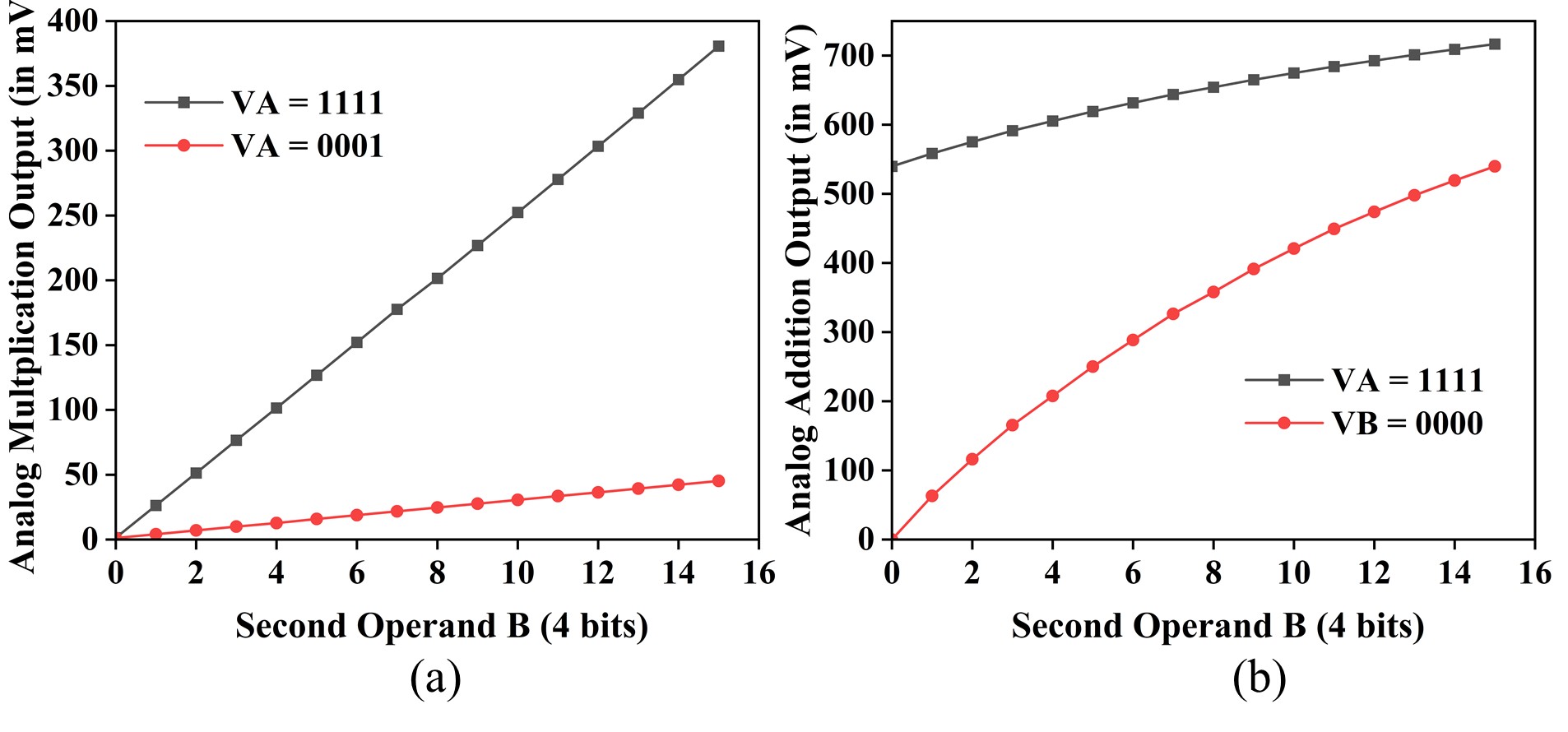}
\vspace{-0.2 in}
\caption{
(a) Analog multiplication output; 
(b) Analog addition output for varying analog voltage levels of operand~A versus different digital levels of operand~B.
}
\label{MUL_ADD_input}
\end{figure}

\begin{figure}[!t]
\centering
\includegraphics[width=1\linewidth]{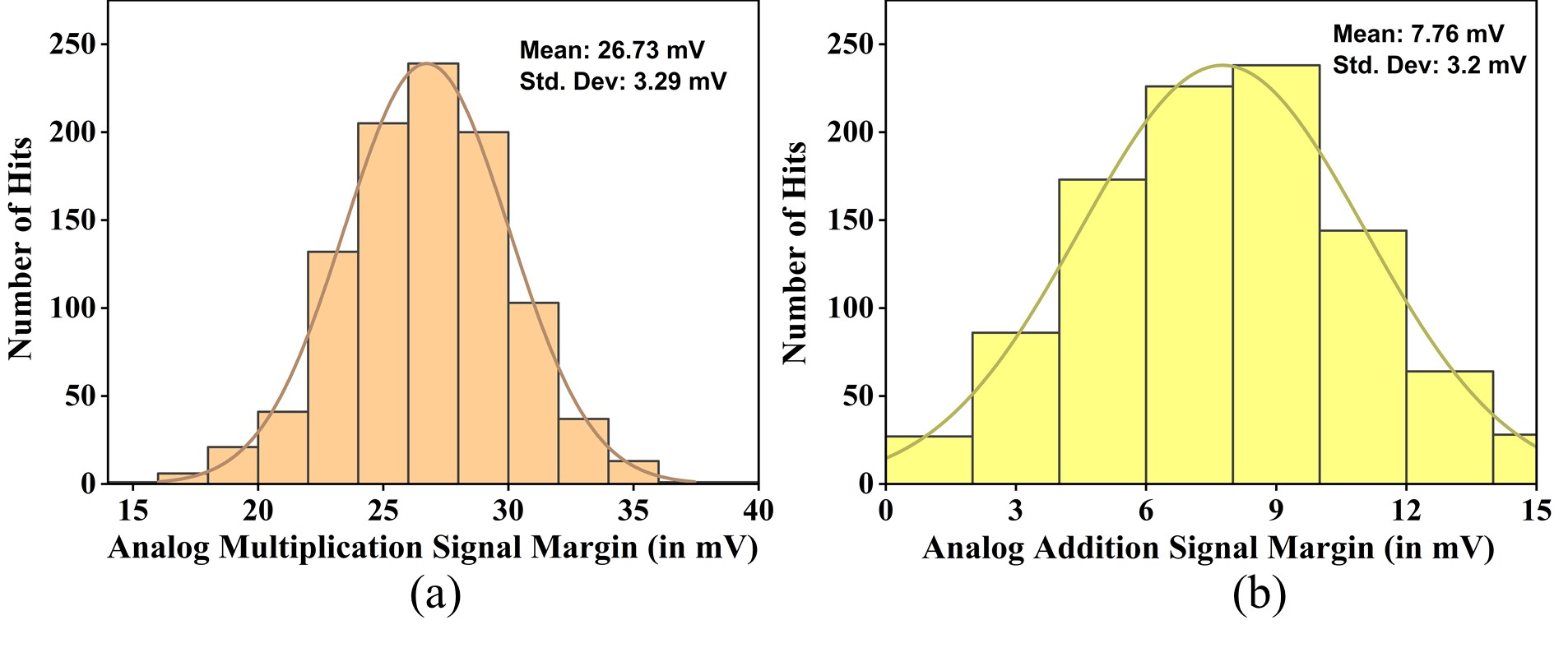}
\vspace{-0.2 in}
\caption{Variation of signal margin for (a) analog multiplication annd (b) analog addition.}
\label{MUL_ADD_MC}
\end{figure}

\begin{figure}[!t]
\centering
\includegraphics[width=1\linewidth]{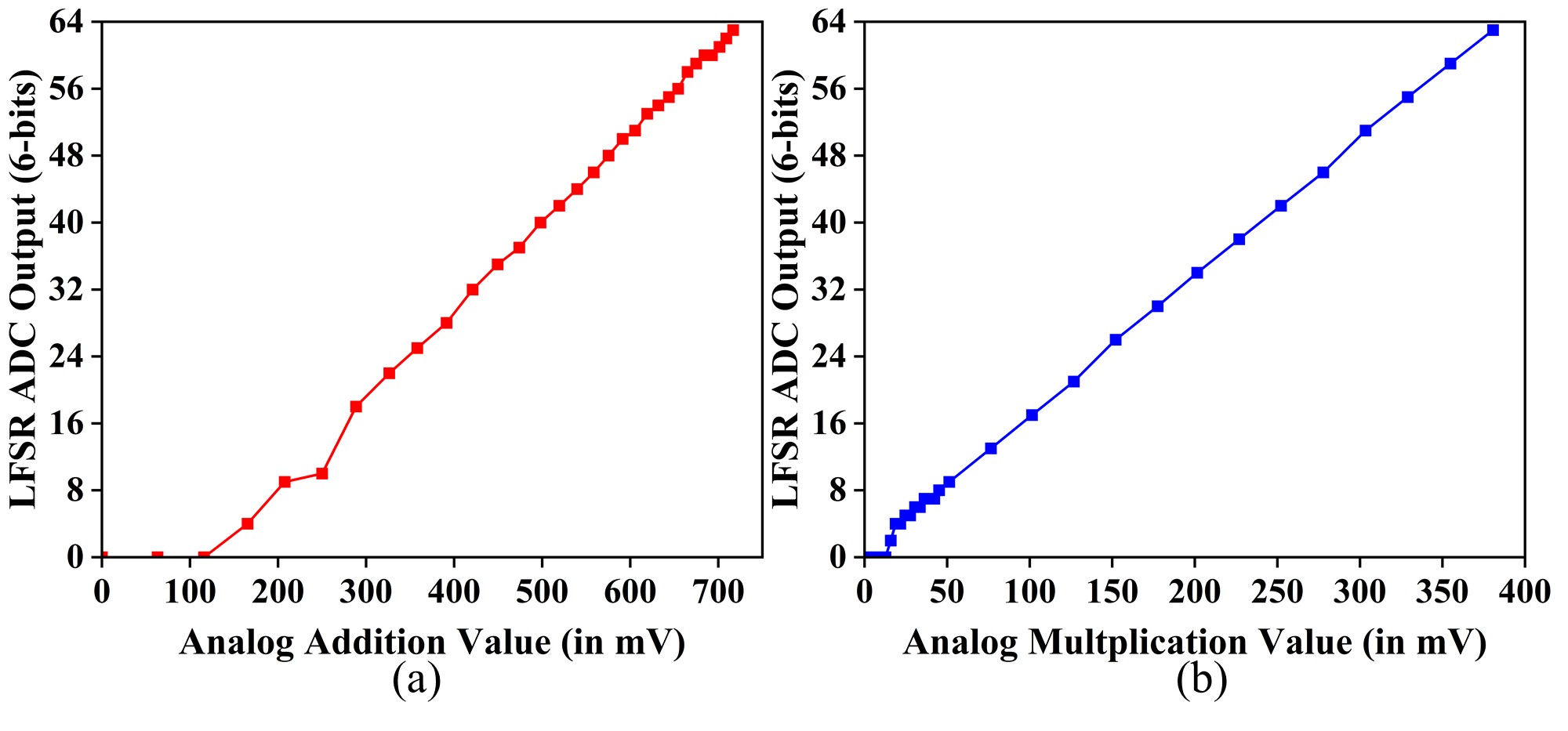}
\vspace{-0.2 in}
\caption{Linearity analysis of 8-bit LFSR ADC for different (a) analog addition and (b) multiplication values.}
\label{LFSR_Linearity}
\end{figure}

\subsection{Element-wise Addition/Multiplication Operation}

The MA-SRAM subarray is employed to generate analog voltages during the multiplication and addition operations. Fig.~\ref{DAC_PC_MC}(a) illustrates the output voltage range of the SRAM-based DAC operating with a 1.8~V supply under different process variations, while Fig.~\ref{DAC_PC_MC}(b) presents the corresponding DAC Signal Margin (SM). The SM of the DAC is defined as the change in the output analog voltage for a 1~LSB variation in the digital input. To generate the analog voltage, the supply is momentarily switched from 0.8~V to 1.8~V for a brief duration of 1~ns. The analog C2C-based multiplier and current-mode adder are implemented to perform multiplication and addition, with their output characteristics shown in Fig.~\ref{MUL_ADD_input}. The signal margins of the analog multiplication and addition operations, obtained from Monte Carlo simulations over 1000 samples, are shown in Fig.~\ref{MUL_ADD_MC}.

For the computation, a PMOS-based differential amplifier is utilized as a comparator in the multiplication operation, while an NMOS-based differential amplifier is used for the addition operation. This design choice is motivated by the respective output voltage ranges of the two operations, the adder output remains near $V_{DD}=0.8$~V, whereas the multiplier output lies closer to ground. To maintain memory subarray density, a simplified differential amplifier structure is adopted, minimizing area overhead at the expense of input-referred offset. To mitigate this offset, an LFSR-based eDRAM calibration phase is performed prior to computation. During calibration, a known input is applied to all the comparator, and the corresponding output code from the LFSR eDRAM is recorded. The expected LFSR output equals the ideal value plus the inherent offset of the differential pair, since standalone LFSR eDRAM simulations without the differential pair under Monte Carlo conditions exhibit negligible intrinsic error. The LFSR output obtained for the known input thus serves as the initial reference point for the LFSR-based ADC operation. This calibration can be performed in parallel, as each word employs an independent comparator with potentially different offsets, leading to distinct initial points. Fig.~\ref{LFSR_Linearity} demonstrates the linearity of the 6-bit digital output equivalent to the 8-bit LFSR output for varying analog addition and multiplication inputs. \textcolor{red}{Based on the Monte Carlo and PVT variation we found the effective number of bits (ENOB) of this LFSR based ADC architecture is 4.78 bits.} 

\begin{figure}[!t]
\centering
\includegraphics[width=1\linewidth]{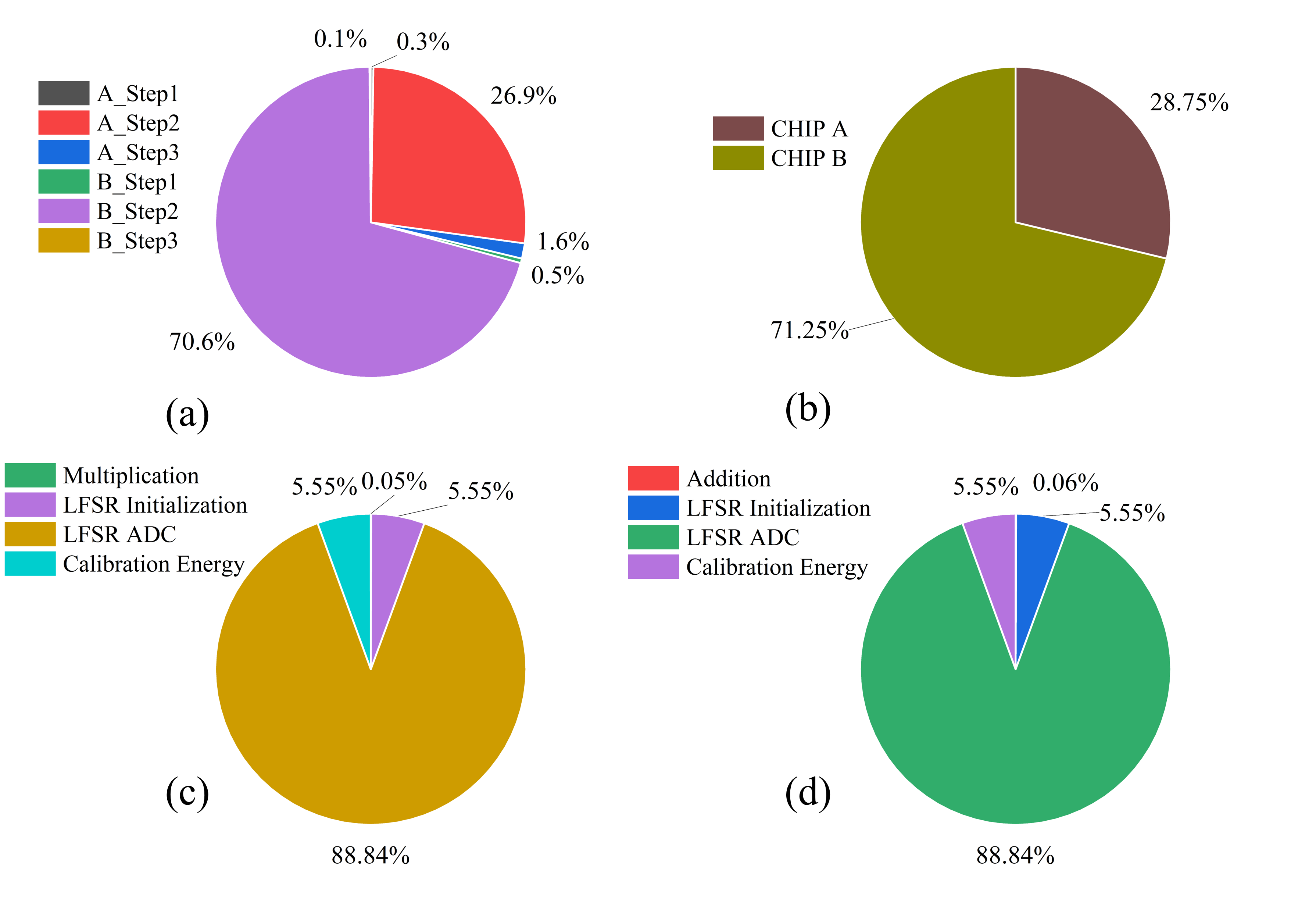}
\caption{(a) Energy breakdown for the transpose operation; (b) energy distribution for the transpose operation between Layer~A (SRAM) and Layer~B (eDRAM); (c) energy breakdown for the element-wise multiplication operation; and (d) energy breakdown for the addition operation.}
\label{pie_energy}
\end{figure}

\subsection{Combined SRAM and eDRAM Bit-cells}

The two proposed SRAM bit-cells (T-SRAM and MA-SRAM) and the corresponding eDRAM bit-cells could, in principle, be combined into a single bit-cell capable of performing transpose, addition, and multiplication operations. However, this approach has not been implemented because it significantly increases circuit complexity and substantially reduces memory density at the word level. Instead, we adopt a more practical design strategy by partitioning the system into separate memory sub-arrays, each tailored for a specific functionality. This functional separation also simplifies 3D integration, leading to a reduced area and a more scalable overall architecture.

\subsection{Latency and Energy Consumption}

The latency and power consumption for various operations were analyzed, accounting for contributions from both the SRAM and eDRAM sub-arrays. For the transpose operation, the total latency depends on the matrix size ($N$), following a latency of $(N + 1)$ cycles. To validate the proposed transpose architecture at a larger scale, we performed parasitic-aware simulations on a $32 \times 32$ matrix with a 4-bit word, requiring a total of $32 \times 128$ memory array. The design achieves a total latency of 264~ns with an 8~ns clock period, and the overall energy consumption is 320.55~nJ. The breakdown of energy consumption for the transpose compute operations, excluding peripheral overhead, is shown in Fig.~\ref{pie_energy}(a) and (b). For 4096 total operations ($32 \times 32 \times 4$), the architecture demonstrates a throughput of 15.51~GOPS and an energy efficiency of 12.77~GOPS/W.

The latency and energy consumption of the addition and multiplication operations consist of multiple contributing components, including the DAC, multiplier/adder, LFSR initial-value write eDRAM power, LFSR-based ADC operations, and calibration energy. The LFSR-based ADC operates over 64 cycles, with each cycle lasting 3~ns for the addition operation and 6~ns for the multiplication operation. The corresponding power breakdown at the word level is shown in Fig.~\ref{pie_energy}(c) and (d). For a $32 \times 32$ matrix with an 8-bit word, the total energy consumption for the multiplication and addition operations is 18.76~nJ and 18.95~nJ, respectively, with latencies of 588~ns and 294~ns. Considering 8192 total operations ($32 \times 32 \times 8$), the resulting throughput and energy efficiency are 13.93~GOPS and 436.61~GOPS/W for multiplication, and 27.86~GOPS and 432.25~GOPS/W for addition, respectively. \textcolor{blue}{Table I presents a comparison of throughput and energy efficiency between the proposed work and prior state-of-the-art in-memory computing designs.}

\begin{table*}[t]
\centering
\caption{\textcolor{red}{Comparison with Previous In-Memory Computing Architectures}}
\label{tab:comparison}
\renewcommand{\arraystretch}{1.25}
\begin{tabular}{|l|c|c|c|c|c|c|c|}
\hline
 & \textbf{Our Work} 
 & \textbf{CIMAT~\cite{cimat}}
 & \textbf{TSRAM~\cite{tsram}}
 & \textbf{CRAM~\cite{cram}} 
 & \textbf{FAT~\cite{fat}} 
 & \textbf{Prop~\cite{prop}}\\
\hline
\textbf{Technology }
& 22\,nm 
& 7\,nm 
& 28\,nm 
& 28\,nm 
& 45\,nm 
& 28\,nm \\

\hline
\textbf{Transpose in situ }
& Yes 
& Yes 
& Yes 
& Yes 
& No 
& No \\

\hline
\textbf{Element-wise Addition }
& Yes 
& No 
& No 
& Yes 
& Yes 
& Yes \\

\hline
\textbf{Element-wise Multiplication }
& Yes 
& No 
& No 
& Yes 
& No 
& Yes \\

\hline
\textbf{Input Precision }
& 4b 
& 8b 
& 4b 
& Arbitrary 
& 4/8b 
& 2--8b \\

\hline
\textbf{Computation Mode }
& Mixed Signal$^{a}$ 
& Analog 
& Analog 
& Digital 
& Digital 
& Digital \\

\hline
\textbf{Foundry Compatible }
& Yes 
& Yes 
& Yes 
& Yes 
& Non-Foundry PDK 
& Yes \\

\hline
\textbf{Parallelism }
& All elements
& Dot-product based
& Dot-product based
& Row-wise 
& Dot-product based
& Column-wise \\

& In parallel
& Row-wise readout
& Row-wise readout
& Readout 
& Readout 
& Readout \\

\hline
\textbf{GOPS Transpose }
& 15.51 
& 3.63$^{c}$ 
& 1.19$^{c}$
& 2.99$^{d}$ 
& -- 
& --
\\

\textbf{GOPS Addition }
& 27.86 
& -- 
& -- 
& 5.73$^{e}$ 
& 29.63 
& 18.08$^{e}$ \\

\textbf{GOPS Multiplication }
& 13.93 
& -- 
& -- 
& 1.11$^{e}$ 
& -- 
& 15.01$^{e}$ \\

\hline
\textbf{GOPS/W Transpose }
& 12.77 
& 0.29-0.32$^{b}$ 
& -- 
& -- 
& -- 
& -- \\

\textbf{GOPS/W Addition }
& 432.25 
& -- 
& -- 
& 229.20 
& -- 
& 7.22$^{e}$ \\

\textbf{GOPS/W Multiplication }
& 436.61 
& -- 
& -- 
& 44.40 
& -- 
& 0.85$^{e}$ \\
\hline
\end{tabular}

\vspace{1mm}
\footnotesize
\begin{flushleft}
$^{a}$ Transpose operation is fully digital, whereas multiplication and addition are performed in the mixed-signal domain.\\
$^{b}$ Estimated GOPS/W at 22\,nm for a $32 \times 32$ array with 4-bit word.\\
$^{c}$ Estimated GOPS at 22\,nm for a $32 \times 32$ array with 4-bit word considering row-wise readout, ADC latency, and memory write latency.\\
$^{d}$ Estimated GOPS at 22\,nm for a $32 \times 32$ array with 4-bit word considering row-wise readout, peripheral latency, and memory write latency at 0.8\,V.\\
$^{e}$ Estimated GOPS or GOPS/W for a $32 \times 32$ array with 8-bit word considering memory write latency.
\end{flushleft}
\end{table*}

\subsection{Area Analysis}

\subsubsection*{Design Rules and Methodology}
 The area analysis presented in this section is based on layouts generated using the GlobalFoundries 22\,nm FDSOI \emph{logic} rule deck. Although memory bit-cells are typically designed using optimized \emph{memory} design rules, all custom bit-cells in this work were implemented using the logic rule set. For reference, a conventional 6T SRAM cell implemented using memory rules occupies approximately $0.1~\mu\text{m}^2$~\cite{gf22sram}. When designed using logic rules at a minimum feature length of 28\,nm, the same 6T cell expands to $0.982~\mu\text{m}^2$ due to stricter spacing and required dummy structures.

\subsubsection*{T-SRAM and T-eDRAM Bit-Cell Areas}
The proposed T-SRAM bit-cell occupies $2.93~\mu\text{m}^2$, which is approximately $2.9\times$ larger than the logic-rule 6T SRAM cell but enables in-memory transpose computation. In contrast, the corresponding T-eDRAM bit-cell is significantly smaller, occupying only $1.04~\mu\text{m}^2$. \textcolor{red}{The 6T SRAM within the T SRAM has the pull-down (WPD), pass-gate (WPG) and pull-up (WPU) transistor sizing ratio set to 2:1:1. For the proposed T-SRAM, the additional write transistor M7 is sized more aggressively for proper write ability, with a width of WM7 = 3.6 × WPD.}

\subsubsection*{Row-Level Area Comparison}
A complete T-eDRAM row with 16 columns and the associated transmission-gate switches occupies $156.37~\mu\text{m}^2$. The equivalent T-SRAM row requires $447.95~\mu\text{m}^2$, primarily due to the larger bit-cell area.

\subsubsection*{MA-eDRAM and MA-SRAM Word Areas}
The MA-eDRAM bit-cell occupies $6.36~\mu\text{m}^2$. An 8-bit MA-eDRAM word, incorporating the multiplier, comparator, digital control logic, and LFSR-based eDRAM, requires a total of $106.43~\mu\text{m}^2$. For the MA-SRAM implementation, a supply voltage of 1.8\,V is required to support digital-to-analog conversion, leading to the use of thick-oxide devices capable of high-voltage operation. The MA-SRAM bit-cell occupies $3.83~\mu\text{m}^2$, while a 4-bit MA-SRAM word requires $44.52~\mu\text{m}^2$.

\subsubsection*{Summary and Discussion}
Although logic-rule-based implementations result in larger bit-cells compared to memory-rule layouts, we adopt logic design rules in this work because they are fully supported by the GlobalFoundries tapeout framework provided by the foundry. T-eDRAM consistently provides the smallest footprint among transpose-capable cells, making it favorable for large crossbars. MA-eDRAM incurs a higher per-bit area due to additional compute circuitry but supports fully in-memory multiplication and addition operations. MA-SRAM offers a smaller bit-cell than MA-eDRAM but requires high-voltage devices. These results illustrate the trade-off between functionality and area: designs supporting richer in-memory operations naturally incur higher area overhead, but they substantially reduce system-level data movement and associated energy costs while enabling general matrix operation within the memory crossbar beyond the conventional CIM dot-product operation.

\section{Conclusion \& Future Direction}
This work presents a 3D-integrated, memory-on-memory SRAM-eDRAM hybrid CIM architecture capable of efficiently performing matrix transpose, element-wise multiplication, addition, and conventional dot-product operations directly within the memory crossbar. Implemented in GlobalFoundries 22~nm FDSOI technology, the proposed design demonstrates the feasibility of extending CIM beyond MAC-centric workloads to support general-purpose matrix computation, thereby broadening the functionality of mixed-signal CIM arrays. By combining specialized sub-arrays with peripheral-aware design principles, the architecture achieves an effective balance among latency, energy efficiency, and compute density. Simulation results show high throughput across multiple matrix kernels, delivering 12.77~GOPS/W for transpose operations and up to 436.61~GOPS/W for arithmetic operations. 

\textcolor{blue}{The proposed 3D stacking of SRAM and eDRAM employs monolithic 3D integration rather than TSV-based or flip-chip bonding approaches. This enables fine-pitch inter-tier connections, supporting high-speed digital-to-analog and analog-to-digital interactions while maintaining high density and low parasitics. The key advantages are threefold.
First, bandwidth enhancement: fine-pitch monolithic inter-tier vias (MIVs) enable massively parallel vertical data transfer, reducing reliance on wide horizontal interconnects and lowering latency compared to 2D or TSV-based designs.
Second, density improvement: stacking SRAM and eDRAM tiers increases effective memory capacity per unit area without expanding the planar footprint, which is particularly beneficial for our proposed macros. Third, power reduction: shorter interconnect lengths and lower parasitic capacitance of MIVs reduce data-movement energy. Monolithic 3D integration also introduces challenges. Thermal coupling between tiers, especially affecting leakage-sensitive SRAM, which can be mitigated through circuit-level design techniques such as power gating and partitioned SRAM sub-arrays, along with architectural redundancy (e.g., spare rows and columns) to enhance robustness. Process variability and device mismatch, aggravated by low-temperature top-tier fabrication, are addressed through word-level computation, array-level averaging, and calibration-aware operation. Finally, we note that the reported area overhead is not fundamental. With continued advances in CMOS fabrication, particularly emerging BEOL-based devices and dense vertical interconnect technologies, the area efficiency of the proposed 3D-integrated architecture can be further improved~\cite{beol1, beol2}.
}

Overall, this memory-on-memory CIM framework generalizes matrix computation in CIM beyond traditional dot-product operations to include element-wise and structural transformations, while still retaining support for standard MAC-based computation. The proposed architecture paves the way toward broader and more versatile CIM use cases in next-generation AI accelerators and general-purpose matrix processing hardware.

\section*{Acknowledgment }
The research was funded in part by National Science Foundation (NSF) through award CCF2319617.

\bibliographystyle{IEEEtran}
\bibliography{references}

\end{document}